\documentclass[useAMS, referee]{biom}

\usepackage{graphicx}
\graphicspath{{./}{figures/}{proj2_app_files/figure-latex/}}
\usepackage{caption}
\usepackage{subcaption}
\usepackage{longtable}
\usepackage{amsmath}
\usepackage{amssymb}
\usepackage{enumerate}
\usepackage{xcolor}
\usepackage{booktabs}
\usepackage{mathptmx}
\usepackage{latexsym}
\usepackage{booktabs}
\usepackage{xltabular}        
\usepackage{threeparttablex}  
\usepackage{ragged2e}
\usepackage{caption}

\newcolumntype{Y}{>{\RaggedRight\arraybackslash}X}   
\newcolumntype{R}{>{\RaggedLeft\arraybackslash}X}    

\title[Model-Assisted Inference for Recurrent Events]{Model-Assisted Causal Inference for the Treatment Effect on Recurrent Events in the Presence of Terminal Events}

\author
{Yiyuan 
Huang$^{1}$, 
Ling 
Zhou$^{2}$, 
Min 
Zhang$^{3}$,
and Peter X.K. 
Song$^{1,*}$\email{pxsong@umich.edu}\\
$^{1}$Department of Biostatistics, School of Public Health, University of Michigan, Ann Arbor, MI 48109, USA \\
$^{2}$Center of Statistical Research, Southwestern University of Finance and Economics, Chengdu, Sichuan, China \\
$^{3}$Vanke School of Public Health, Tsinghua University, Beijing, China 
}

\let\bibhang\relax
\usepackage{natbib}

\begin{document}

\pagerange{\pageref{firstpage}--\pageref{lastpage}} 

\label{firstpage}

\begin{abstract}
 This paper is motivated by evaluating the benefits of patients receiving mechanical circulatory support (MCS) devices in end-stage heart failure management inference, in which hypothesis testing for a treatment effect on the risk of recurrent events is challenged in the presence of terminal events. Existing methods based on cumulative frequency unreasonably disadvantage longer survivors as they tend to experience more recurrent events. The While-Alive-based (WA) test has provided a solution to address this survival-length-bias problem, and it performs well when the recurrent event rate holds constant over time. However, if such a constant-rate assumption is violated, the WA test can exhibit an inflated type I error and inaccurate estimation of treatment effects. To fill this methodological gap, we propose a Proportional Rate Marginal Structural Model-assisted Test (PR-MSMaT) in the causal inference framework of separable treatment effects for recurrent and terminal events. Using the simulation study, we demonstrate that our PR-MSMaT can properly control type I error while gaining power comparable to the WA test under time-varying recurrent event rates. We employ PR-MSMaT to compare different MCS devices with the postoperative risk of gastrointestinal bleeding among patients enrolled in the Interagency Registry of Mechanically Assisted Circulatory Support program. 
\end{abstract}


\maketitle




\section{Introduction}
\label{sec:intro}

Heart transplantation remains the gold standard for patients with end-stage heart failure, but its use is restrained from the severe shortage of donor organs, with only about 3,000 adult transplants performed annually in the United States. Durable mechanical circulatory support (MCS) has therefore become a choice of heart failure management in the current clinical practice, with over 2,500 operations of ventricular assist device implants reported each year in the Interagency Registry for Mechanically Assisted Circulatory Support (INTERMACS) since year 2006. Such data first motivated our methodology development and were then analyzed in this paper. In effect, patients who receive continuous-flow left ventricular assist devices (LVADs) are dominant in implants, whereas biventricular assist devices (BiVADs) are less popular and only used for patients who have severe right ventricular dysfunction. Although these devices have markedly improved survival and quality of life, they are accompanied by a substantial burden of adverse events, including bleeding, infection, stroke, and device malfunction. Among these, gastrointestinal bleeding (GIB) is one of the most frequent and clinically consequential complications, often necessitating repeated hospitalizations, transfusions, and endoscopic procedures. The recurrence of GIB is known to be associated with increased mortality and diminished quality of life \citep{hammer2024gastrointestinal}. One important clinical question of central interest is whether, for clinically comparable patients who would survive for the same duration, the choice between BiVAD and LVAD device implants leads to a different risk of recurrent GIB events.

Despite its clinical importance, this question remains unresolved for two main reasons. First, single-center studies generally lack adequate sample sizes of device users to provide sufficiently powered statistical comparisons. Second, randomized clinical trials are logistically difficult and ethically complex; patients and clinicians may be reluctant to randomize device type due to uncertainty about comparative risks. The published evidence in the current literature has thus far been inconclusive on the benefit of the two devices. Some studies suggest that early planned BiVAD support improves some clinical outcomes relative to delayed LVAD-to-BiVAD conversion \citep{Fitzpatrick2009}, whereas others report higher bleeding and infection rates among BiVAD recipients \citep{Cleveland2011}. Large-scale registry data such as INTERMACS may offer a promising source of data to deliver more convincing conclusions to the question. Of note, INTERMACS aggregates patient-level data from hospitals nationwide, and preliminary findings have uncovered lower average survival for BiVAD patients \citep{Kirklin2015}. It is worth pointing out that prior analyses cannot disentangle whether differences in bleeding reflect true device effects or underlying patient survival differences, which are the major confounding factors. The differential survival distribution between the two device groups complicates the analysis and interpretations. This motivates us to develop rigorous, interpretable statistical methods to obtain valid data evidence from large electronic health record (EHR) databases to address the above challenges.

Our investigation focuses primarily on the risk of recurrent GIB, a leading post-surgical complication that substantially contributes to morbidity, cost, and resource utilization. Because GIB can occur multiple times during post-surgery analytic follow-up, recurrent event analysis is the method of choice for this outcome \citep{andersen1982cox, pepe1993some, lawless1995some}. One major challenge in the use of this analytic pertains to the presence of terminal events such as deaths, which precludes future recurrences and induces a competing risk scenario. Simply treating death as a type of censoring yields inference conditional on surviving subjects, which may bias comparisons between devices with patients' different survival profiles and suffer from the lack of causal interpretability.

Several statistical approaches have been proposed to address survival bias in recurrent event analyses. The Ghosh–Lin estimator \citep{ghosh2000nonparametric} incorporates death into the estimation of the marginal mean number of recurrent events, but it could penalize groups with longer survival by accumulating more events. To mitigate this, while-alive (WA) estimands are introduced to normalize the mean number of events by the restricted mean survival time, providing a measure of event rate per unit time alive \citep{mao2023nonparametric, schmidli2021estimands, wei2023properties}. While this WA test provides an important advance, it implicitly assumes that the recurrent event rate is constant over time. In practice, event intensities often evolve, such as rising early after surgery and tapering with time, which makes the WA test prone to having type I error inflation when the baseline rate is time-varying or when survival distributions differ substantially between device groups. These limitations call for new methods that can accommodate both time-varying recurrent-event rates and differential survival distributions across treatment groups.

The concerns described above are highly relevant in our motivating example. First, we notice that the type of ventricular assist device may influence GIB recurrence through two distinct pathways: a direct effect on the recurrent bleeding process (e.g., via shear stress and anticoagulation requirements) and an indirect effect mediated through survival, since longer-lived patients have more time at risk. Disentangling these two pathways is crucial for valid inference on the device-specific bleeding risk. In alignment with this clinical insight, we adopt the separable effects framework, proposed by \citet{stensrud2022separable} and extended to recurrent–terminal event settings by \citet{janvin2023causal}, which enables us to decompose the treatment (i.e., device) into two hypothetical components acting separately on the recurrent and terminal pathways; see Figure \ref{fig:r5} with detailed explanations in Section \ref{sec3: se}. In short, this decomposition yields an interpretable estimand for the direct effect of treatment on recurrent events, not restricted to survivors, and accommodates time-varying baseline rates. Our contribution in this paper is a new statistical inference method that enriches the existing work, which has primarily focused on estimation under the separable effects framework, with limited development of formal hypothesis tests.

Our development of statistical inference is devoted to a model-assisted score-type hypothesis test, termed PR-MSMaT, which aims to test the direct treatment effects on recurrent events in the presence of terminal events. Our approach builds on the separable effects framework to produce a single, time-integrated test with a clear causal mechanistic interpretation. As shown in the paper, PR-MSMaT can control type I error well under varying baseline event rates and differential survival, overcoming the limitations of the WA-based test. 

The paper is organized as follows. Section 2 introduces a hypothesis testing procedure in the WA paradigm, where we demonstrate the limitations of the WA-based test by a simulation experiment. Section 3 introduces the Separable Effects Framework (SEF) \citep{stensrud2022separable, janvin2023causal}, which is adopted in our development of a hypothesis test procedure PR-MSMaT. Section 4 develops the Proportional Rate Marginal Structural Model-assisted Test (PR-MSMaT) and discusses its analytic properties. Section 5 focuses on simulation experiments to compare the proposed PR-MSMaT with the WA–based test in terms of type I error and statistical power in various scenarios. Section 6 illustrates the utility of the proposed methods to the INTERMACS dataset to test the difference in GIB burden between ventricular support devices, followed by some concluding remarks in Section 7. 

\section{While-Alive-based (WA) Test: A Revisit}
\label{sec2: wa}

Methods for analyzing recurrent events in the presence of a terminal event have evolved considerably over the past decades. The Nelson--Aalen (NA) estimator was among the earliest approaches, where death was treated as a censoring event for the recurrence process. This assumption implicitly allows individuals to continue accruing events after death, leading to systematic overestimation of the recurrent event burden in settings with non-negligible mortality. To address this issue, acknowledging no further event after the survival time $D$, \cite{ghosh2000nonparametric} (GL) proposed a nonparametric framework based on the mean frequency function, $\mu(t) = \mathbb{E}\{N^{*}(t)\}$, where $N^{*}(t)$ is the cumulative number of recurrent events up to time $t$. \cite{ghosh2000nonparametric} also developed the corresponding hypothesis test procedure. However, the GL test does not properly account for the reality among patients receiving device implants; that is, the device type that prolongs survival may increase recurrent event counts simply because patients live longer, not necessarily because the device elevates morbidity.

To address this limitation, \cite{mao2023nonparametric} considered an approach based on the while-alive (WA) estimand, which averages the number of cumulative recurrent events over survival time. Specifically, at a prespecified time $\tau$, the While-Alive Loss Rate (WALR) is given by 
$$
\ell(\tau) = \frac{\mathbb{E}\{N^{*}(\tau)\}}{\mathbb{E}(D \wedge \tau)}.
$$
This quantity represents the expected number of recurrent events per unit time in life over the interval $[0,\tau]$, so it corrects for the survival-induced bias inherent in both NA and GL methods. \cite{mao2023nonparametric} further derived an inference for treatment comparisons with WALR, in which a modified estimand $L(t)=\ell(t)t$ is further considered to avoid the unstable behavior of $\ell(t)$ for small $t$, termed {\it the survival-completed cumulative loss function}. 

It is worth pointing out that the test based on the WA estimand only incurs a fair comparison of the recurrent event if the true baseline recurrent event rate is constant over time. Arguably, this condition may not hold in many practical studies. For instance, in our motivating example, it is irrational to believe that patients' GIB risk trajectory remains constant over time. In such cases, the existing WA-based test may become invalid with inflated type I errors. 

\begin{figure}[htbp]
    \centering
    \includegraphics[width=\linewidth]{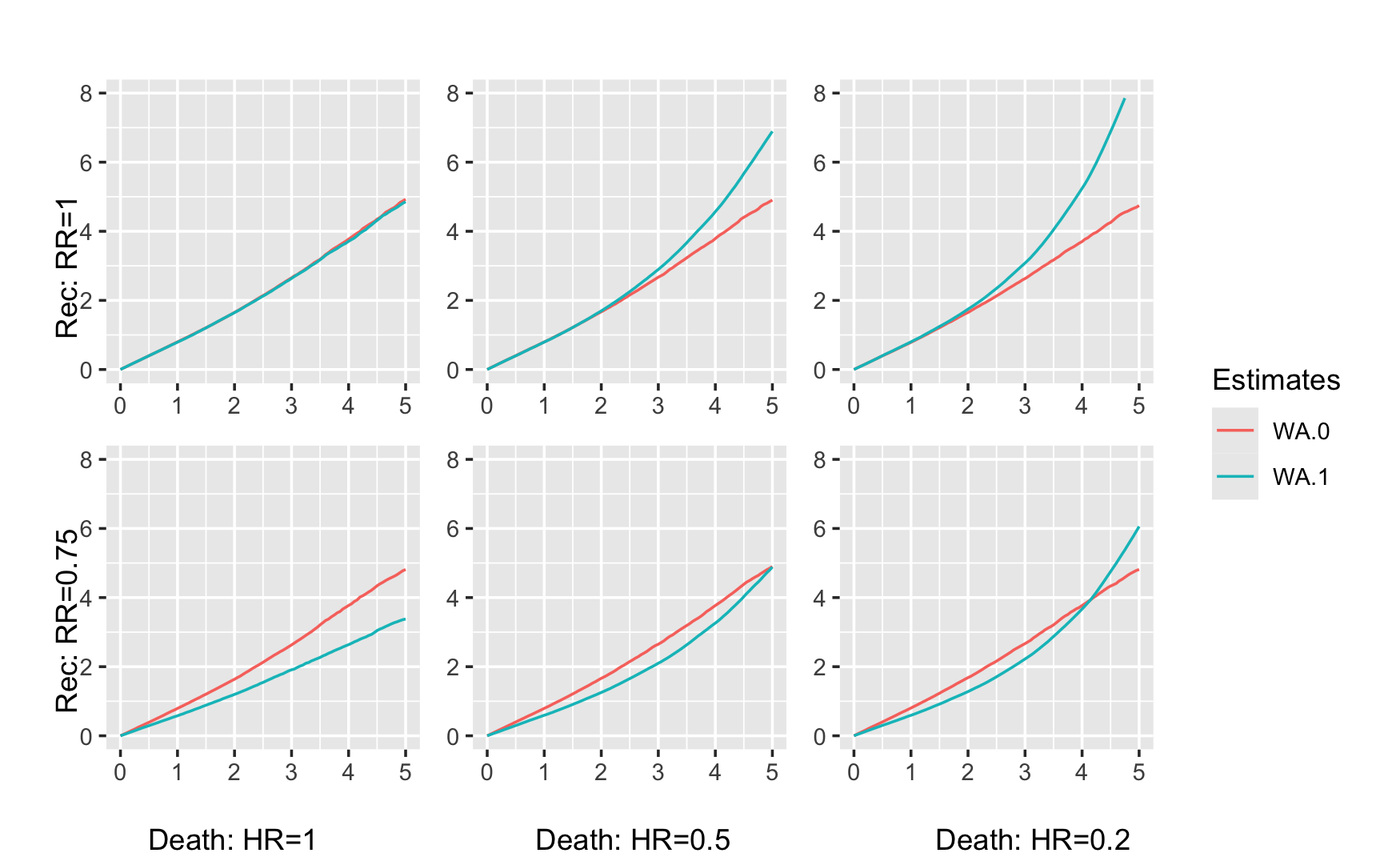}
    \caption{The simulation results illustrate the limitations of the While-alive-based test. WA.0: the survival-completed cumulative loss function for the control group; WA.1: the survival-completed cumulative loss function for the treatment group; RR: Recurrent event Ratio; HR: Hazard Ratio.}
    \label{fig:r1}
\end{figure}

To illustrate this potential pitfall, we conducted a simulation experiment, emulating a randomized trial under time-varying recurrent event rates. For simplicity, we assumed independent recurrent events and independence between recurrent and terminal events. The baseline recurrent event rate was set as a time-varying stepwise function, starting at $2$ events per year and then rising by a factor of $\exp(0.5)$ every 1.2 years within a maximum follow-up of 5 years. Treatment effects were specified through two recurrent event rate ratios (RR $=1.00, 0.75$) together with three terminal event hazard ratios (HR $=1.0, 0.5, 0.2$), yielding six scenarios. Each scenario was simulated with $n=500$ subjects and 100 rounds of Monte Carlo replications, and the WA function (survival-completed cumulative loss function) was estimated for treatment and control arms using an existing R package {\it WA} \citep{mao2023nonparametric}. Figure~\ref{fig:r1} displays the WA functions averaged over 100 replications. From which, we learned: 
\begin{enumerate}[(i)]
    \item In those scenarios where the treatment has no impact on recurrent events (RR $=1$) but does affect survival (HR $\neq1$), such as the scenario of RR $=1.0$ and HR $=0.5$, the WA estimates appeared substantially different between the two treatment groups. Consequently, a hypothesis test based on such WA estimates could have a poor type I error control.
    \item In the settings where the treatment can truly affect both recurrent and terminal events, for example, the scenario of RR $=0.75$ and HR $=0.2$, the estimated WA curves crossed at year 4 or so (i.e., not really separable), which implied that the WA-based test could lose statistical power to test the treatment effect around year 4. 
\end{enumerate}

Such numerical evidence illustrates that with time-varying baseline event rates, the WA method may have inflated type I error and lose statistical power for hypothesis testing when the treatment acts through both the recurrent process and survival. This calls for the need for a new inference approach that can explicitly consider the pathways linking treatment, recurrent events, and terminal events.

\section{Separable Effects Framework (SEF)}
\label{sec3: se}

\begin{figure}
    \centering
    \includegraphics[width=\textwidth]{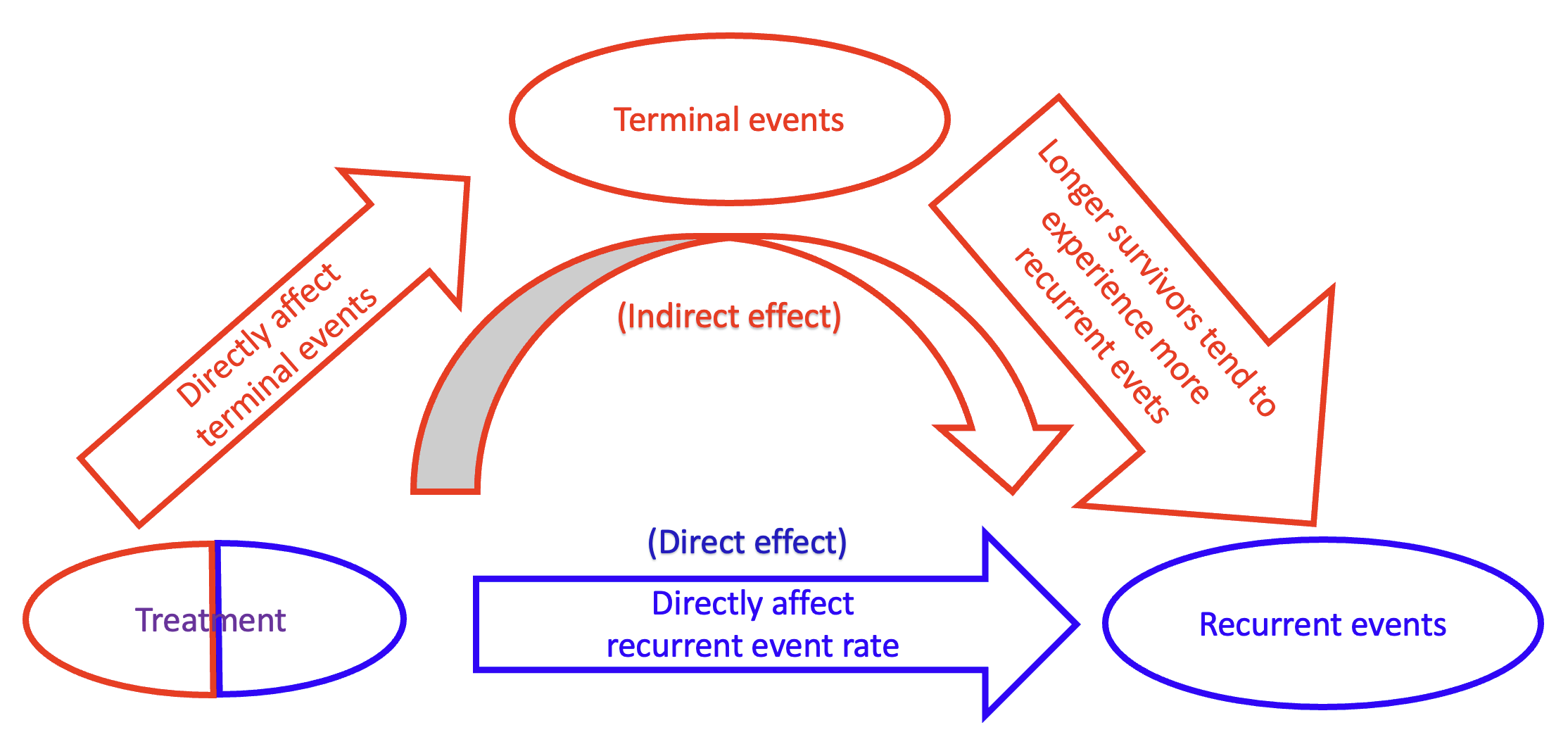}
    \caption{Intuition regarding the relationship between the treatment, recurrent events, and terminal events.}
    \label{fig:r5}
\end{figure}

To better evaluate treatment effects on recurrent events, it is useful to consider two pathways: as shown in Figure~\ref{fig:r5}, a {\it \textbf{direct effect}}, where a treatment directly affects on the recurrent process not through survival, such as GIB in our motivating example, and an {\it \textbf{indirect effect}}, where a treatment influences the recurrent process through survival since longer survivors tend to experience more recurrent event. The primary goal is to disentangle these two pathways and isolate the direct effect in the evaluation, while excluding the indirect survival-mediated effect \citep{stensrud2022separable}. 

The Separable Effects Framework (SEF) was previously introduced by \cite{stensrud2022separable} for binary outcomes with competing events and later extended by \cite{janvin2023causal} to recurrent event settings. Several other extensions were also proposed \citep{stensrud2021generalized, stensrud2023conditional}. Adopting this framework, we aim to develop a hypothesis testing procedure for the direct effect in the SEF paradigm. In this paper, we mainly consider a randomized trial setting, in which we focus on the relationships between the treatment, recurrent process, and terminal events, ignoring all the other confounders. In cases where the treatment was not randomized, the statistical techniques (such as matching or IPTW weighting) may be used to balance the treatment groups in order to apply this SEF method. 

\begin{figure}[ht]
    \centering
    \begin{subfigure}[t]{0.33\textwidth}
        \includegraphics[width=\linewidth]{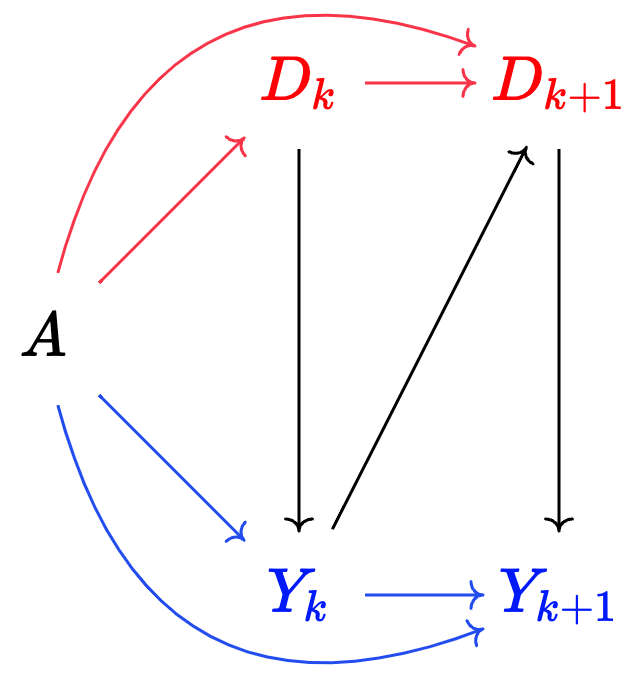}
        \caption{}
        \label{fig:r3}
    \end{subfigure}
    \hfill
    \begin{subfigure}[t]{0.45\textwidth}
        \includegraphics[width=\linewidth]{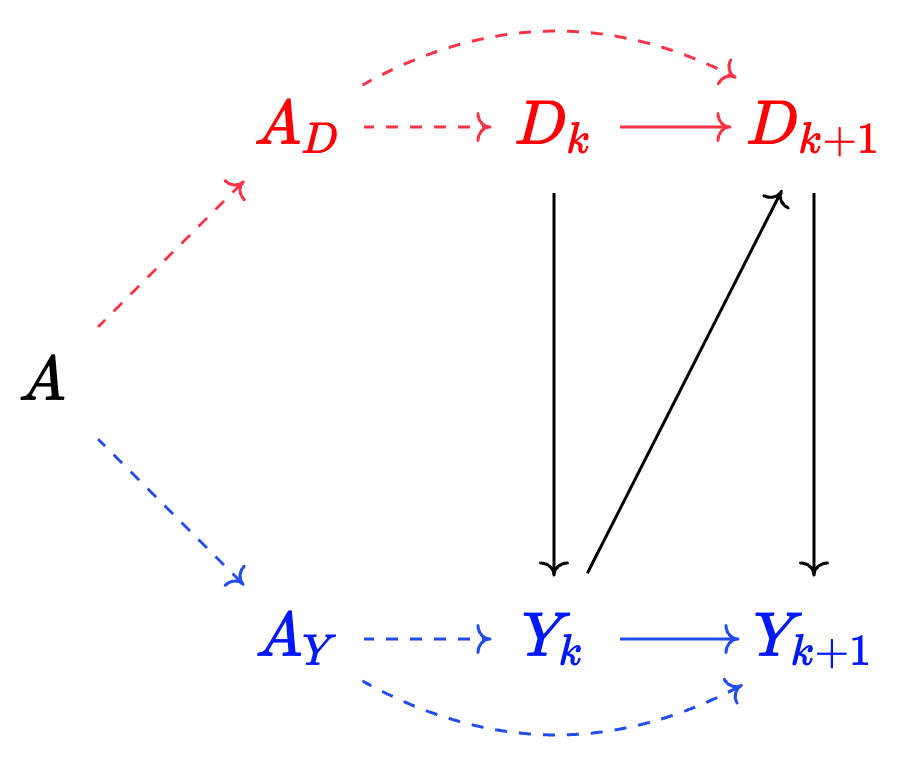}
        \caption{}
        \label{fig:r4}
    \end{subfigure}
    \caption{Directed acyclic diagram (DAG) illustrates:
    (a) the causal pathways for treatment $A$, recurrent event $Y_k$, and terminal event $D_k$; 
    (b) the hypothetical decomposition of treatment $A$ into $A_Y$ (directly affecting recurrent events) and $A_D$ (directly affecting terminal events).}
    \label{fig:r3r4}
\end{figure}

Let $A \in \{0,1\}$ denote the primary treatment under investigation. Consider a discrete-time partition of the follow-up time window in the form $0 = t_0 < t_1 < \cdots < t_K = \tau$, where $\tau$ is the maximum follow-up time. For an interval cell $(t_{k-1},t_k]$, let $D_k$ be the death indicator within the time interval, and $Y_k$ be the cumulative number of recurrent events by $t_k$. Without loss of generality, we define the topological order in a hypothesized Directed Acyclic Graph (DAG) shown in Figure~\ref{fig:r3} within the time interval $(t_{k-1},t_k]$, with $D_{k}$ occurring before $Y_k$. For convenience, for an arbitrary process $X_k$, we denote its change by $\Delta X_k := X_k-X_{k-1}$, and the history by $\bar{X}_k = \left\{X_j\right\}_{j=0}^k$.

In the SEF paradigm, the treatment $A$ is supposed to be decomposed into two distinct treatment components: $A_Y$, which solely influences the occurrence of recurrent events, and $A_D$, which exclusively impacts terminal events, such that receiving $A_{Y} = A_{D} = a$ could result in the same outcomes as receiving $A = a$, $a\in\{0,1\}$. See Figure~\ref{fig:r4} for the organization of these two components in the SEF framework. Although in practice, we cannot always identify the exact decomposition with two treatment components, conceptually we may leverage such a hypothetical framework to conduct a causal inference for their mechanistic pathways. 

Formally, define counterfactual recurrent outcomes as $Y_k^{a_{Y}, a_{D}}$, representing the cumulative number of recurrent events by $t_k$ had the subject received $A_{Y}=a_{Y}$ and $A_{D}=a_{D}$. Similarly, we define the counterfactual death indicators $D_k^{a_{Y},a_{D}}$. The counterfactual mean function is then given by 
$$
\mu_k^{a_{Y},a_{D}} = E\left[ Y_k^{a_{Y},a_{D}} \right].
$$ 
It follows that the \textbf{separable direct effect} is captured by contrasts of $\mu_k^{a_{Y}=1,a_{D}}$ versus $\mu_k^{a_{Y}=0,a_{D}}$ for fixed $a_{D}$, while the \textbf{separable indirect effect} by contrasts of $\mu_k^{a_{Y},a_{D}=1}$ versus $\mu_k^{a_{Y},a_{D}=0}$ for fixed $a_{Y}$ (Janvin et al., 2023). Such definitions conceptually partition the overall treatment effect into distinct and interpretable pathways as shown in Figure~\ref{fig:r5}.

To identify separable effects, we impose the following standard assumptions of causal inference: 
\begin{enumerate}[(i)]
    \item Consistency: If $A=a$, then $\bar D_{K}=\bar D_{K}^a$, $\bar Y_{K}=\bar Y_{K}^a$;
    \item Exchangeability: $(\bar Y^a_{K}, \bar D^a_{K}) \perp A$;
    \item Positivity: $f_{\bar D_{k+1}, Y_k}(0,y_k)>0 \text{ implies } P(A=a|\bar D_{k+1}=0, \bar Y_k= y_k)>0$, for all $a\in \left\{0, 1\right\}$, $k \in \left\{0, \cdots, K\right\}$. 

    Additionally, for the SEF, the following conditions hold:
    \item Dismissible component conditions: 
    $$
    \begin{aligned}
        Y_{k+1}&\perp A_D|A_Y, \bar D_{k+1}, \bar Y_k;\\
        D_{k+1}&\perp A_Y|A_D, \bar Y_{k}, \bar D_k, \text{for all } k\in \left\{0, \cdots, K\right\}. 
    \end{aligned}
    $$
\end{enumerate}

The conditions in (iv) ensure that the decomposed $A_{Y}$ and $A_{D}$ stay at non-overlapping pathways. Under the identification conditions, the G-formula for the counterfactual mean function is 
$$
\mu_k^{a_{Y},a_{D}} = \sum_{t=1}^k E\left[ \frac{I(A=a_{Y})}{\pi_A(a_{Y})} \prod_{j=0}^t \frac{\pi^{{a_{D}}}_{D_j}}{\pi^{a_{Y}}_{D_j}} \, \Delta Y_t \right],\;\; k\in \left\{0, \cdots, K\right\}, 
$$ 
where $\pi_A(\bullet)=P(A=\bullet)$ and $\pi^\bullet_{D_j}=f_j^\bullet(D_j)$, with $f_j^\bullet(\circ)=P(D_j=\circ|\bar Y_{j-1}, \bar D_{j-1}, A=\bullet)$. 

For each time interval, the change in the mean function is estimated by 
\begin{equation}
\label{eq1}
    \Delta \hat\mu_t^{a_{Y},a_{D}} = \hat S_t^{a_{Y}} \Delta \hat B_t^Y,
\end{equation}
where $\hat S_t^{a_{Y}}$ is the discrete Kaplan–Meier estimator of the survival function for group $a_{Y}$ and 
$$
\Delta \hat B_t^Y = \frac{\sum_{i=1}^n \hat W^D_{i,t}(a_{Y},a_{D}) I(A_i=a_{Y}) Z_{i,t} \Delta Y_{i,t}}{\sum_{j=1}^n I(A_j=a_{Y}) Z_{j,t}},
$$
where $Z_{i,t}$ is the at-risk indicator and 
$$
\hat W^D_{i, t}(a_Y, a_D) = \prod_{q=1}^{t}\left( \frac{1-\Delta \hat \Lambda_{i,q}^{D|\mathcal{F}}(a_D)}{1-\Delta \hat \Lambda_{i, q}^{D|\mathcal{F}}(a_Y)}\right)^{1-D_{i,q}} \left( \frac{\Delta \hat \Lambda_{i, q}^{D|\mathcal{F}}(a_D)}{\Delta \hat \Lambda_{i, q}^{D|\mathcal{F}}(a_Y)} \right)^{D_{i, q}},
$$
with $\Delta\hat\Lambda_{i, q}^{D|\mathcal{F}}(a)=\hat P(D_{i, q}=1|\bar D_{i,q-1}=0, \bar Y_{i, q-1}, A_i=a)$. The cumulative estimator is then $\hat\mu^{a_{Y},a_{D}}_k = \sum_{t=1}^k \Delta \hat\mu^{a_{Y},a_{D}}_t$ \citep{janvin2023causal}.

The weighting process $\hat W^D_{i, t}(a_Y, a_D)$ attempts to mimic a hypothetical randomized trial on which $A_Y$ and $A_D$ may be implemented separately. Specifically, it adjusts the observed data distribution so that the counterfactual survival process under $a_D$ is comparable to that under $a_Y$, taking effect only when $a_D \neq a_Y$. Intuitively, this calibration via the weights is used to eliminate the imbalance between groups even the treatment $A$ is randomized, allowing us to recover the distribution of recurrent events had $A_Y$ and $A_D$ been independently randomized. The weights scale down those individuals who are more likely to die under the counterfactual $a_D$ than under their observed assignment, and scale up those who are less likely to die, effectively reshaping the risk set to reflect the distribution of recurrent events under the counterfactual intervention. In this way, the weighting aligns the survival experience across groups so that differences in the estimated recurrent event process can be attributed solely to the direct effect of $A_Y$, rather than being biased by differential survival induced by $A_D$.

The above review of the SEF introduces a framework in which the treatment can be rigorously decomposed into two components acting exclusively on recurrent and terminal events. The direct effect in the SEF, therefore, represents the contrast in the recurrent event process attributable solely to $A_Y$, while holding $A_D$ fixed. By separating this key pathway of interest from the indirect effect, we can test hypotheses about the direct treatment effects on recurrent events while mitigating the distortions observed with the WA method under time-varying recurrent event rates.

\section{Proportional Rate Marginal Structural Model-assisted Test (PR-MSMaT)}
\label{sec4: PR-MSMaT}

Building on the SEF, we develop a hypothesis test for the direct treatment effect on recurrent events. The logic is analogous to conventional survival analysis: in a similar spirit of the Kaplan–Meier estimator motivating both the log-rank test and the Cox proportional hazards model with treatment as the sole covariate, the separable estimators proposed by \cite{janvin2023causal} motivate a Proportional Rate Marginal structural model (PR-MSM) along the lines of a score-based test.

Let $\mu^{a_Y,a_D}(t)$ denote the counterfactual marginal mean number of recurrent events by time $t$ had the treatment components been set at $(a_Y,a_D)$. The PR-MSM takes the form:
\begin{equation}
\label{PR-MSM}
    \mu^{a_Y,a_D}(t)=\mu^{a_Y=0,a_D}(t) \exp\big(\beta^{a_D} a_Y\big), \quad a_D \in \{0,1\},
\end{equation}
suggesting that, for each fixed $a_D$, the counterfactual mean functions across $a_Y$ differ by a constant multiplicative factor over time. It follows that $\beta^{a_D}$ can be estimated by solving the following estimation equation:
\begin{equation}
\label{eq2}
\int_0^{\tau} \sum_{i=1}^n Z_i(t)A_i \Big\{ d\hat{\mu}^{a_Y=1,a_D}(t) - \exp(\beta^{a_D})\, d\hat{\mu}^{a_Y=0,a_D}(t) \Big\} = 0,
\end{equation}
where $\hat{\mu}^{a_Y,a_D}(t)$ is the counterfactual mean function estimator and $Z_i(t)$ is the at-risk process. See Appendix I for the detailed derivation for (\ref{eq2}).

Under $H_0: \beta^{a_D}=0$ (no direct treatment effect), the left-hand side of (\ref{eq2}) reduces to 
\begin{equation}
\label{eq3}
    U_n \;=\; \int_0^\tau \sum_{i=1}^n Z_i(t) A_i \Big\{ d\hat{\mu}^{a_Y=1,a_D}(t) - d\hat{\mu}^{a_Y=0,a_D}(t) \Big\}.
\end{equation} 
Let $\widehat{\mathrm{Var}}(U_n)$ denote a consistent estimator of $\mathrm{Var}(U_n)$, which can be estimated by the transforming cumulative hazard estimates based on differential equations \citep{ryalen2018transforming}. Our PR-MSM-assisted Test (PR-MSMaT) statistic is formed as the standardized score test statistic:
$$
T_n \;=\; \frac{U_n}{\widehat{\mathrm{Var}}(U_n)^{1/2}}.
$$

\begin{theorem}
\label{thm:PR-MSMaT}
Suppose that (i) $\hat{W}_{i,t}^D(a_Y,a_D)$ and $\hat{S}_{t}^{a_Y}$ are uniformly consistent to their population parameters; (ii) the risk process $Z_i(t)$ is bounded away from 0 and $\infty$; and (iii) the standard martingale Lindeberg conditions hold. Then, under $H_0:\beta^{a_D}=0$, the test statistic $T_n \;\overset{d}{\longrightarrow}\; N(0,1)$ as $n\to\infty$.
\end{theorem}

Theorem \ref{thm:PR-MSMaT} shows that the proposed PR-MSMaT is a valid test for the null hypothesis of no separable direct treatment effect on recurrent events. See Appendix II for the proof of Theorem \ref{thm:PR-MSMaT}.

The PR-MSMaT extends the separable effect estimators by providing a formal hypothesis test that summarizes treatment effects across the pre-specified follow-up time period. Unlike the original estimators, which yield time-specific contrasts that may be harder to synthesize, the proposed test delivers an overall inference for the direct treatment effect. It enjoys interpretability for broader audiences in clinical practice; in particular, it provides the causal interpretability within the SEF paradigm, in a similar way to the log-rank or Cox score test widely used in practice. Simulation studies in Section~\ref{sec: simu} show that our PR-MSMaT maintains a proper type I error control and achieves higher power than while-alive-based tests in scenarios where recurrent event rates vary over time, a realistic situation encountered often in practice, including our motivating example.

\section{Simulation Experiment}
\label{sec: simu}

Following the simulation design given in \cite{janvin2023causal}, we evaluated the finite-sample performance of our PR-MSMaT compared with the WA-based test under a 2-arm randomized trial for treatment $A$, with treatment group $A=1$ and the control group $A=0$. In this case, all contrasts are evaluated by treatment versus control. Outcomes were generated under an additive intensity model over $K=1000$ intervals $(t_{k-1}, t_k]$, $k=1,\dots,1000$. For each interval, we consider 
$$
\begin{aligned}
    P(\Delta Y_k = 1 \mid A_Y, D_k=0) &= \beta_{Y,0,k} + \beta_{Y,A} A_Y,\\
    P(D_k = 1 \mid A_D, \bar Y_{k-1}) &= \beta_{D,0} + \beta_{D,A} A_D + \beta_{Y,D} Y_{k-1},
\end{aligned}
$$
with at most one recurrent event per interval. We fix $\beta_{D,A}=-0.5/K$, set the null case with $\beta_{Y,A}=0$ to evaluate type I error scenarios, and $\beta_{Y,A}=-0.5/K$ for power evaluation. Each configuration is examined via $N=1000$ rounds of Monte Carlo replicates with sample size $n=1000$. The three types of baseline recurrent-event rate $\beta_{Y,0,k}$ are designed according to the following temporal patterns: 
\begin{enumerate}[(a)]
    \item \textbf{Constant:} $\beta_{Y,0,k}=2/K$ for all $k$.
    \item \textbf{Decreasing:} Starting at $3/K$ and declining by $0.5/K$ every 200 intervals (down to $1/K$). 
    \item \textbf{Increasing:} Starting at $1/K$ and rising by $0.5/K$ every 200 intervals (up to $3/K$).
\end{enumerate} 
We examine three types of hypotheses: \textbf{Two-sided:} $H_0:\theta=0$ vs. $H_1:\theta\neq 0$; \textbf{Left-sided:} $H_0:\theta \geq 0$ vs. $H_1:\theta < 0$; \textbf{Right-sided:} $H_0:\theta \leq 0$ vs. $H_1:\theta > 0$. The parameter $\theta$ represents the difference in recurrent events between treatment and control groups, which refers to $\beta^{a_D}$ for our PR-MSMaT and the WALR ratio for the WA-based test. 

\begin{figure}[htbp]
  \centering
  \includegraphics[width=\linewidth]{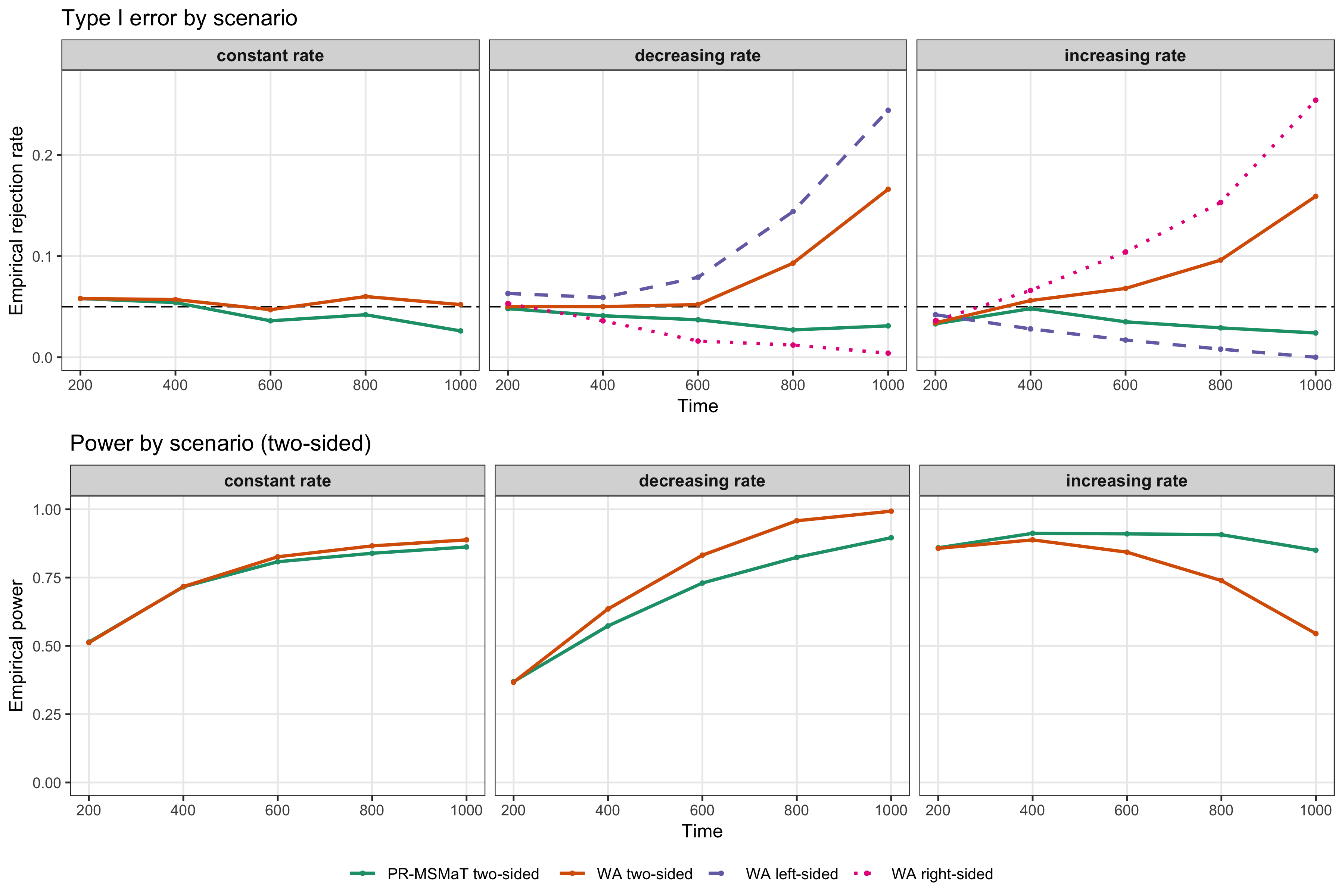}
\caption{Empirical type I error rate and power across time under three scenarios designed in the simulation study: constant, decreasing, and increasing baseline recurrent-event rates. The dashed horizontal line marks the nominal significance level 0.05.}
  \label{fig:type1-power}
\end{figure}

Simulation results are summarized in Figure~\ref{fig:type1-power}. When the baseline recurrent event rate is constant (i.e., scenario (a)), both our PR-MSMaT and the WA test maintain a nominal type I error rate and achieve similar power. This agreement under time-homogeneous rates confirms that both methods are well-behaved in this scenario. In contrast, when the baseline recurrent event rate varies with time, their performances diverge. Under a decreasing rate (i.e, scenario (b)), WA exhibits inflated type I errors for both the two-sided test and the left-sided test. Under an increasing rate (i.e., scenario (c)), WA again shows inflated type I errors for both the two-sided test and the right-sided test. Amazingly, our PR-MSMaT clearly controls type I errors well in all scenarios. In terms of power, our PR-MSMaT achieves comparable power to the WA test in scenario (b) and even larger power than WA in scenario (c). Note that, in both scenarios (b) and (c), WA has the disadvantage of inflated type I errors. This demonstrates that PR-MSMaT performs reliably against nonstationary baseline recurrent-event processes and clearly outperforms the WA test.

Directional test results help explain the WA behavior. The WA test summarizes the data by scaling the cumulative number of recurrent events by the restricted mean survival time, in which the “while-alive rates” are compared across arms. This procedure implicitly treats the observed total event count of each subject as if it arose from a roughly constant per-unit-time rate over that subject’s observed survival. When the true baseline rate decreases over time, most events happen earlier, so a subject who dies earlier tends to have a larger fraction of their lifetime spent in the high-rate period. In this case, WA imputes a higher “average event rate while alive” to that shorter-lived profile, even if the subject is not actually more event-prone than someone who lives longer. As a result, WA can systematically take earlier deaths as evidence of a higher event intensity in the control arm; this pushes the estimated treatment effect (treatment vs. control) downward and inflates left-sided type I error. Likewise, when the baseline rate increases over time, events concentrate later, so longer survivors accumulate more events simply because they remain under observation during the high-rate period. WA handles it as an arm-specific elevation in the “while-alive rate”, biasing the estimated treatment effect upward and inflating right-sided type I error.

By contrast, our PR-MSMaT directly targets the separable, pathway-specific effect of treatment on the recurrent event process, holding the survival pathway fixed. Because it does not attribute differences driven purely by unequal follow-up windows in high- or low-rate portions of the hazard curve to a treatment effect, our PR-MSMaT avoids these directional biases. As a result, it preserves nominal type I error across all time-varying baseline patterns while still delivering competitive power.

\section{Application}
\label{sec:app}

We now analyze the data of the motivating example introduced in Section~\ref{sec:intro}. Within the SEF, the question of interest is whether device type has a significant direct effect on recurrent gastrointestinal bleeding (GIB) when the survival pathway is held fixed. In this formulation, any difference in recurrent event rates can be attributed to the device itself rather than to differences in survival.
Our dataset contains adults (age $\geq$ 19 years) who underwent a primary continuous-flow LVAD implant, with or without subsequent RVAD (BiVAD), between April 1, 2006 and December 31, 2017. We excluded pulsatile devices, isolated RVAD, total artificial hearts, and non-primary LVAD implants. The primary outcome is a recurrent event GIB, defined as bleeding from the upper or lower GI tract or occult-positive events of unknown source. Death is treated as a terminal (or nuisance) event. To emulate the randomized trial setting, we performed a 1:1 propensity-score matching using pre-implant demographic, clinical, laboratory, echocardiographic, and hemodynamic factors. Baseline characteristics were well balanced post-match (all $p$-values for univariate tests $>0.05$). See Supplementary materials for the summary tables for baseline factors in the dataset after matching. After 1:1 matching, the analytic sample comprises $n=1{,}051$ patients per group (BiVAD and LVAD). 

\begin{figure}[htbp]
\centering
\includegraphics[width=0.48\linewidth]{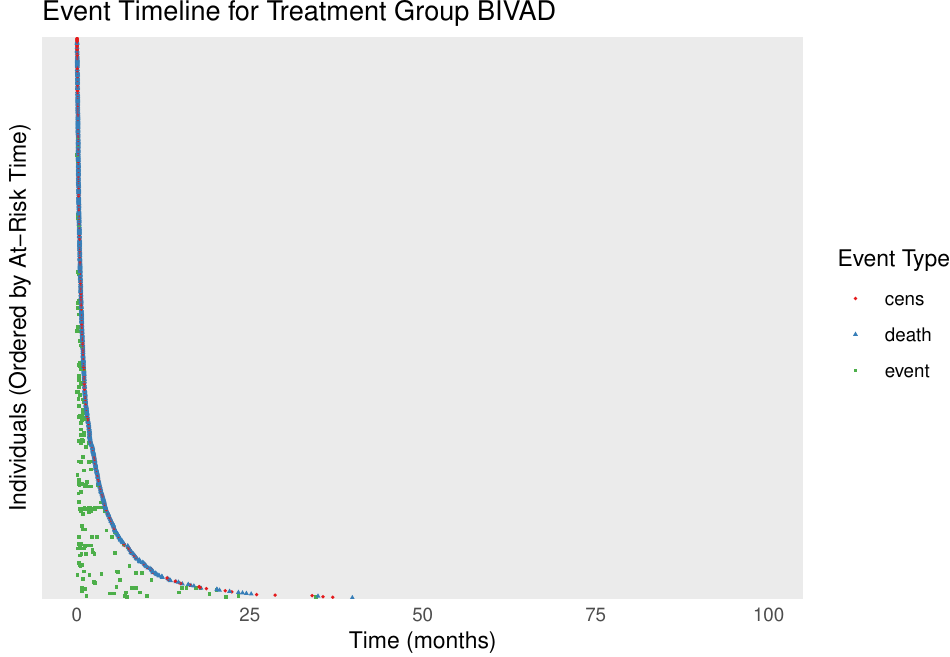}
\hfill
\includegraphics[width=0.48\linewidth]{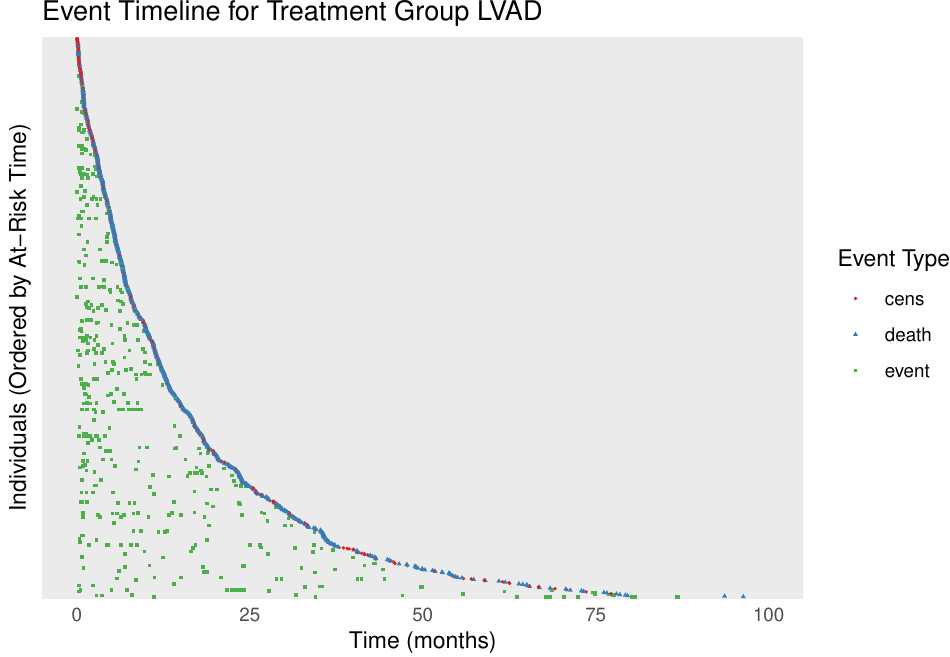}
\caption{Plots of times for recurrent event of gastrointestinal bleeding (GIB) marked by green dots, death by blue dots appearing at the edge, and censoring by red dots among the matched pairs across device type (LVAD vs.\ BiVAD).}
\label{fig:app-gib-summary}
\end{figure}

\begin{figure}[htbp]
    \centering
    \includegraphics[width=\textwidth]{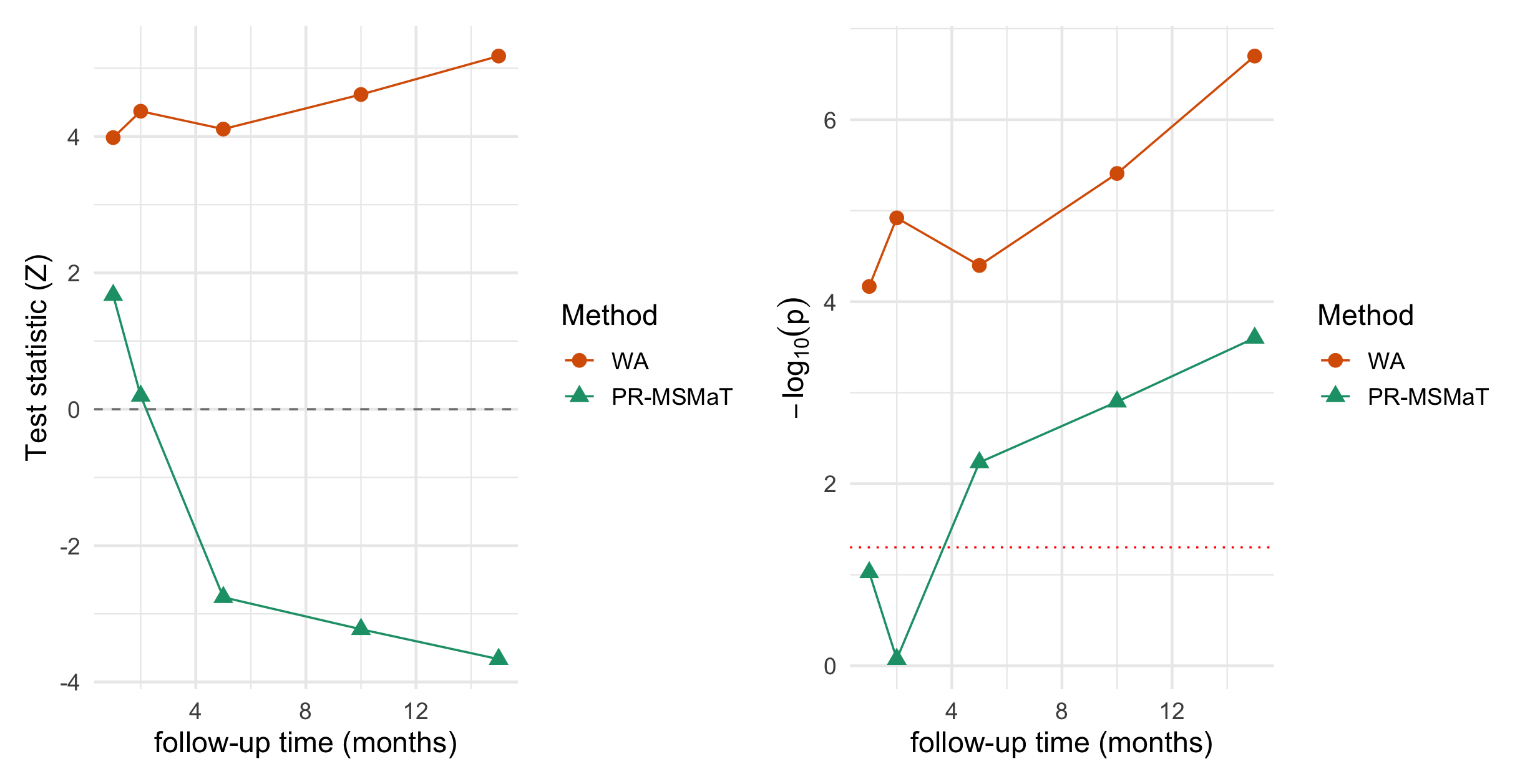}
    \caption{Comparison of test statistics ($Z$) and their $-\log_{10}(p)$ values at five selected follow-up times (months) across two test methods: existing WA (While-alive-based test) and new PR-MSMaT (Proportional Rate Marginal Structural Model–assisted Test). The horizontal dashed line in the left panel indicates $Z=0$, and the red dotted line in the right panel corresponds to the conventional significance threshold $p=0.05$.}
    \label{fig:app-results}
\end{figure}

We summarized recurrent gastrointestinal bleeding (GIB) events and overall survival in the matched cohort by device type (LVAD vs.\ BiVAD). Figure~\ref{fig:app-gib-summary} displays two descriptive plots for GIB burden as well as their follow-up time. In the survival comparison as seen clearly, the BiVAD group experienced markedly earlier mortality than the LVAD group. We also evaluated the treatment contrast between BiVAD and LVAD using the WA test and our PR-MSMaT at 5 time horizons $\tau\in\{1,2,5,10,15\}$ (months). Figure~\ref{fig:app-results} reports the standardized $Z$ values and $p$-values, where both methods use the contrast $\text{BiVAD}-\text{LVAD}$, implying that negative $Z$ favors BiVAD.

It is noticeable that WA and PR-MSMaT in fact target different estimands, so they react differently to the survival imbalance in Figure~\ref{fig:app-gib-summary} where people in the BiVAD group tend to die much earlier. WA divides the survival-completed cumulative frequency by the mean survival time, which reduces survivorship bias but does not fully remove it. If the event rate is time varying, especially higher soon after implant, earlier mortality in BiVAD means a larger fraction of its observed time lies in the high-risk window, which can inflate the while-alive rate and yield a positive $Z$, indicating more severe GIB burden in the BiVAD group, even without a higher direct propensity for events in that group. PR-MSMaT uses separable-effects weighting to align the survival pathway $(A_D)$ across groups and tests the direct effect on recurrent events $(A_Y)$ while holding the death process fixed. In Figure~\ref{fig:app-results}, the negative PR-MSMaT $Z$ values at larger follow-up time ($\tau$) indicate that, after fixing the survival pathway, BiVAD has a lower direct recurrent-event burden (fewer GIB events attributable to the direct effect) than LVAD. Thus, the sign disagreement (WA positive vs.\ PR-MSMaT negative) reflects an indirect survival pathway that over-weights high-risk early periods for BiVAD.

Hospitals participating in the Society of Thoracic Surgeons (STS) INTERMACS leverage standardized data forms to collect longitudinal data on this population. Use of data for this study was approved by the STS Participant User File Task Force. Patient consent to submit data to STS INTERMACS is not required by STS and is determined by local hospital policy. The opinions expressed in this manuscript do not represent those of STS.

\section{Discussion}

We developed an inference method, termed PR-MSMaT, a score-type test that is constructed under the separable effects framework (SEF) to evaluate the direct treatment effect on recurrent events while explicitly holding the survival pathway fixed. Conceptually, PR-MSMaT evaluates the separable effect estimands of Stensrud et al. (2022) and Janvin et al. (2023) by a practical score test analogous to a log-rank test or Cox score test. It targets a pathway-specific contrast rather than a mixture of mechanisms driven by both recurrent and terminal processes, aggregating information over the entire follow-up to yield a single, interpretable summary statistic with a well-defined causal estimand. In extensive simulations across diverse data-generating regimes, PR-MSMaT has been shown to maintain nominal type I error and demonstrated robustness to time-varying baseline hazards. Notably, it preserves validity in scenarios where the existing while-alive (WA) test exhibits systematic type I error inflation. In effect, the SEF effectively decouples recurrent-event dynamics from differential survival and reduces spurious associations induced by varying survival lengths.

In our empirical study, we observe that fixing the survival pathway enables us to isolate the component of recurrent gastrointestinal bleeding (GIB) burden attributable solely to the device choice itself, independent of mortality differences. Such finding is clinically relevant; it may help make clinical decisions on device selection to minimize bleeding risk, conditional on patients' expected survival. Furthermore, because the SEF provides causal interpretability under a well-executed 1:1 matching scheme, our comparison between two types of devices shows appealing insights useful to improve patients' post-implant quality of life. Technically, our method PR-MSMaT allows the time horizon $\tau$ and baseline rate function to vary, an important flexibility to analyze clinical data collected in other settings. These features position PR-MSMaT as a useful, well-justified analysis tool to evaluate treatments beyond implant devices in either randomized trials or registry-based observational studies.

Several directions warrant further development of PR-MSMaT. Extending the procedure to simultaneously accommodate multiple recurrent endpoints (e.g., bleeding, infection, and device malfunction) and center-level heterogeneity would enhance its utility for multicenter registries and composite safety monitoring. Developing formal sensitivity analyses for potential violations of the dismissible component conditions would strengthen its application to observational data, where pathway isolation may be imperfect. Finally, incorporating covariate-adaptive weighting and sample-size formulas calibrated to pathway-specific alternatives could improve trial efficiency and power, particularly for registry-embedded or platform studies in which evolving practice patterns interact with event dynamics.

\section*{Acknowledgments}

The authors acknowledge the use of data from the Society of Thoracic Surgeons. This research was partially supported by NIH R01ES033656 grant (Song). 

%


\bibliographystyle{biom}
\bibliography{reference_use}   

\section*{Supporting Information}
A summary table referenced in Section~\ref{sec:app} is available
with this paper at the Biometrics website on Wiley Online Library. This table shows satisfactory 1:1 matching results for the baseline factors in the data to justify the application of the proposed causal inference in the empirical study. 

\vspace*{-8pt}

\appendix

\section{I. Derivation of Estimating Equation for $\beta^{a_D}$ in PR-MSM}
\label{ap.i}

By rewriting the equation (\ref{eq1}) in Section~\ref{sec3: se} in continuous-time format and denote the recurrent event process by the counting process notation $N(t)$ \citep{fleming2013counting}, the separable increment estimator is
\[
 d\hat{\mu}^{a_Y,a_D}(t)
 = \frac{\sum_{i=1}^n Z_i(t)\, I(A_i=a_Y)\, \hat{S}^{a_Y}(t)\, \hat{W}_i^{D}(a_Y,a_D; t)\, dN_i(t)}{\sum_{j=1}^n Z_j(t)\, I(A_j=a_Y)}.
\]
Specializing to $a_Y=1$ gives
\begin{equation}
 d\hat{\mu}^{a_Y=1,a_D}(t)
 = \frac{\sum_{i=1}^n Z_i(t) A_i\, \hat{S}^{a_Y=1}(t)\, \hat{W}_i^{D}(1,a_D; t)\, dN_i(t)}{D_1(t)},
 \tag{A.1}
\end{equation}
where
\[
 D_1(t) := \sum_{j=1}^n Z_j(t) A_j.
\]

The PR-MSM in equation (\ref{PR-MSM}) posits
\[
 \mu^{a_Y=1,a_D}(t)=\mu^{a_Y=0,a_D}(t)\exp(\beta^{a_D})
 \quad\Longleftrightarrow\quad
 d\mu^{a_Y=1,a_D}(t)=\exp(\beta^{a_D})\, d\mu^{a_Y=0,a_D}(t).
\]
To connect the observed data to the model for counterfactual means $\mu^{a_Y,a_D}(t)$, we replace each $d\mu^{a_Y,a_D}(t)$ by its consistent estimator $d\hat{\mu}^{a_Y,a_D}(t)$ \citep{janvin2023causal}:
\begin{equation}
    d\hat\mu^{a_Y=1,a_D}(t)=\exp(\beta^{a_D})\, d\hat\mu^{a_Y=0,a_D}(t).
    \tag{A.2}
\end{equation}

Plugging (A.2) on the left of (A.1) while keeping the right-hand side as the estimator yields the moment condition:
\begin{equation}
 \exp(\beta^{a_D})\, d\hat\mu^{a_Y=0,a_D}(t)
 = \frac{\sum_{i=1}^n Z_i(t) A_i\, \hat{S}^{a_Y=1}(t)\, \hat{W}_i^{D}(1,a_D; t)\, dN_i(t)}{D_1(t)}.
 \tag{A.3}
\end{equation}

Multiplying both sides of (A.3) by $D_1(t)$ and rearranging leads to
\begin{equation}
 \sum_{i=1}^n Z_i(t) A_i\, \hat{S}^{a_Y=1}(t)\, \hat{W}_i^{D}(1,a_D; t)\, dN_i(t)
 \; -\; \exp(\beta^{a_D})\, D_1(t)\, d\hat\mu^{a_Y=0,a_D}(t)\;=\;0.
 \tag{A.4}
\end{equation}
Now replace the unknown $d\mu^{a_Y=0,a_D}(t)$ by its estimator, obtained from the same equation (\ref{eq1}) in Section~\ref{sec3: se} with $a_Y=0$, we have
\begin{equation}
    d\hat{\mu}^{a_Y=0,a_D}(t)
 = \frac{\sum_{i=1}^n Z_i(t) (1-A_i)\, \hat{S}^{a_Y=0}(t)\, \hat{W}_i^{D}(0,a_D; t)\, dN_i(t)}{D_0(t)},
 \tag{A.5}
\end{equation}
where
\[
 D_0(t):=\sum_{j=1}^n Z_j(t) (1-A_j).
\]
Substituting (A.5) into (A.4) gives the estimating equation with combined summations on both treatment groups:
\begin{align*}
 &\sum_{i=1}^n Z_i(t) A_i\, \hat{S}^{a_Y=1}(t)\, \hat{W}_i^{D}(1,a_D; t)\, dN_i(t)
 \; \\
 -&\; \exp(\beta^{a_D})\, \frac{D_1(t)}{D_0(t)}\, \sum_{i=1}^n Z_i(t) (1-A_i)\, \hat{S}^{a_Y=0}(t)\, \hat{W}_i^{D}(0,a_D; t)\, dN_i(t)
 \;=\;0.
\end{align*}
Integrating $t$ over $[0,\tau]$ yields the sample estimating equation for $\beta^{a_D}$:
\begin{align*}
&\int_0^{\tau}\Bigg[\sum_{i=1}^n Z_i(t) A_i\, \hat{S}^{a_Y=1}(t)\, \hat{W}_i^{D}(1,a_D; t)\, dN_i(t) \\
 &- \exp(\beta^{a_D})\, \frac{D_1(t)}{D_0(t)}\,
   \sum_{i=1}^n Z_i(t) (1-A_i)\, \hat{S}^{a_Y=0}(t)\, \hat{W}_i^{D}(0,a_D; t)\, dN_i(t)\Bigg]=0.
 \tag{A.6}
\end{align*}

Finally, noting that the first summation corresponds to $d\hat{\mu}^{a_Y=1,a_D}(t)$ and the second to $d\hat{\mu}^{a_Y=0,a_D}(t)$, we can write (A.6) compactly as
$$
 \int_0^{\tau} \sum_{j=1}^n Z_j(t)A_j \Big\{ d\hat{\mu}^{a_Y=1,a_D}(t) - \exp(\beta^{a_D}) d\hat{\mu}^{a_Y=0,a_D}(t) \Big\} = 0.
$$

\section{II. Proof of Theorem~\ref{thm:PR-MSMaT} (Asymptotic normality of PR-MSMaT)}

Let $N_i(t)$ denote the recurrent-event counting process for subject $i$ on $[0,\tau]$,
$Z_i(t)$ the at-risk indicator, and $A_i\in\{0,1\}$ the randomized treatment.
Write $M_i(t)=N_i(t)-\Lambda_i(t)$, where $\Lambda_i(t)$ is the compensator of $N_i(t)$
with respect to the natural filtration. Then $\{M_i(t)\}$ is a square-integrable local 
martingale with predictable variation $\langle M_i\rangle(t)$ 
\citep{andersen2012statistical}.

For $g\in\{0,1\}$, define 
$D_g(t)=\sum_{j=1}^n Z_j(t)\,I(A_j=g)$ and assume 
$D_g(t)/n\to p_g(t)$ uniformly with $\inf_{t\le\tau}p_g(t)>0$.

Let $\hat S^{a_Y}(t)$ be a uniformly consistent estimator of $S^{a_Y}(t)$ and 
$\hat W_i^D(a_Y,a_D;t)$ uniformly consistent for the corresponding weights.
Define the separable mean-increment estimator
\[
d\hat\mu^{a_Y,a_D}(t)
=\frac{\sum_{i=1}^n Z_i(t)\,I(A_i=a_Y)\,\hat S^{a_Y}(t)\,
\hat W_i^D(a_Y,a_D;t)\,dN_i(t)}{D_{a_Y}(t)},
\]
which corresponds to (\ref{eq1}) in Section~\ref{sec3: se}.

From (\ref{eq3}) in Section~\ref{sec4: PR-MSMaT}, the PR-MSMaT score under $H_0\!:\beta^{a_D}=0$ is
\begin{equation}
U_n=\int_0^\tau \sum_{j=1}^n Z_j(t)A_j
\Big\{d\hat\mu^{1,a_D}(t)-d\hat\mu^{0,a_D}(t)\Big\}.
\tag{A.7}
\end{equation}

Fix $a_Y\in\{0,1\}$. Writing
\[
d\hat\mu^{a_Y,a_D}(t)
=\frac{1}{D_{a_Y}(t)}\sum_{i=1}^n I(A_i=a_Y)\,Z_i(t)\,
\hat S^{a_Y}(t)\,\hat W_i^D(a_Y,a_D;t)\,dN_i(t),
\]
replace $dN_i(t)$ by $dM_i(t)+d\Lambda_i(t)$ and expand the numerator and denominator 
around their expectations. A standard Taylor expansion for ratios of empirical processes,
combined with uniform consistency of $\hat S^{a_Y}$ and $\hat W_i^D$ and the uniform LLN
for predictable bounded processes \citep{andersen2012statistical}, yields
\begin{equation}
d\hat\mu^{a_Y,a_D}(t)
=\mu'^{\,a_Y,a_D}(t)\,dt 
+\frac{1}{n}\sum_{i=1}^n \phi_i^{a_Y,a_D}(t)\,dM_i(t)
+r_n^{a_Y,a_D}(t),
\tag{A.8}
\end{equation}
with $\sup_{t\le\tau}|r_n^{a_Y,a_D}(t)|=o_p(n^{-1/2})$, and
\[
\phi_i^{a_Y,a_D}(t)
=\frac{n\,I(A_i=a_Y)\,Z_i(t)\,S^{a_Y}(t)\,W_i^D(a_Y,a_D;t)}{D_{a_Y}(t)}.
\]

Substituting (A.8) into (A.7),
\[
U_n
=\frac{1}{n}\sum_{i=1}^n \int_0^\tau \psi_i(t)\,dM_i(t) + o_p(n^{1/2}),
\]
where $\psi_i(t)=\sum_{j=1}^n Z_j(t)A_j\{\phi_i^{1,a_D}(t)-\phi_i^{0,a_D}(t)\}$.

Define
\[
\xi_{n,i}:=\int_0^\tau \frac{\psi_i(t)}{n^{3/2}}\,dM_i(t),
\quad\text{so that}\quad
\sum_{i=1}^n \xi_{n,i}=n^{-1/2}U_n+o_p(1).
\]

The predictable variation satisfies
\begin{equation*}
V_n=\sum_{i=1}^n\langle\xi_{n,i}\rangle
=\sum_{i=1}^n \int_0^\tau \frac{\psi_i^2(t)}{n^3}\,d\langle M_i\rangle(t).
\end{equation*}

Assume:  
(i) i.i.d.\ subjects;  
(ii) bounded predictable $\phi_i^{a_Y,a_D}(t)$;  
(iii) $\inf_t D_g(t)/n>0$ and $\sup_t D_g(t)/n<\infty$;  
(iv) the martingale Lindeberg condition \citep{andersen2012statistical, hall2014martingale}.  

Because increments are scaled by $1/n^{3/2}$ and weights are bounded, (iv) holds automatically.
Thus $V_n\xrightarrow{p}\sigma^2\in(0,\infty)$.

By Rebolledo’s martingale central limit theorem 
\citep{rebolledo1980central},
\[
\sum_{i=1}^n \xi_{n,i}\xrightarrow{d}N(0,\sigma^2),
\quad\text{equivalently}\quad
n^{-1/2}U_n\xrightarrow{d}N(0,\sigma^2).
\]

A consistent variance estimator is
\[
\hat\sigma^2
=\sum_{i=1}^n \int_0^\tau \frac{\{\hat\psi_i(t)\}^2}{n^3}\,dN_i(t),
\]
so that $\widehat{\mathrm{Var}}(U_n)=n\,\hat\sigma^2$ satisfies
\[
\frac{\widehat{\mathrm{Var}}(U_n)}{n}\xrightarrow{p}\sigma^2.
\]

Hence the PR-MSMaT statistic
\[
T_n
=\frac{U_n}{\{\widehat{\mathrm{Var}}(U_n)\}^{1/2}}
=\frac{n^{-1/2}U_n}{\hat\sigma}
\xrightarrow{d}N(0,1),
\]
under $H_0:\beta^{a_D}=0$. 

\end{document}